\documentclass[apj]{emulateapj}
\usepackage{amsmath}
\newcommand \degree{^{\circ}}
\def\mean#1{\overline{#1}}
\def\dim#1{\mbox{\,#1}}
\newcommand \chimps{\ h^{-1} \dim{Mpc}}

\begin{document}


\title[Measuring Gas Accretion and Angular Momentum near Simulated SMBHs]{Measuring Gas Accretion and Angular Momentum near Simulated Supermassive Black Holes}
  
\author{Robyn Levine\altaffilmark{1}}
\author{Nickolay Y. Gnedin\altaffilmark{2,3,4}}
\author{Andrew J. S. Hamilton\altaffilmark{5,6}}

\altaffiltext{1}{CITA, 60 St. George St., Toronto, ON, M5S 3H8,Canada; 
levine@cita.utoronto.ca} 
\altaffiltext{2}{Particle Astrophysics Center, Fermi National Accelerator Laboratory, Batavia, IL 60510, USA} 
\altaffiltext{3}{Kavli Institute for Cosmological Physics, The University of Chicago, Chicago, IL 60637, USA}
\altaffiltext{4}{Department of Astronomy \& Astrophysics, The University of Chicago, Chicago, IL 60637, USA}
\altaffiltext{5}{JILA, University of Colorado, Boulder, CO 80309, USA}
\altaffiltext{6}{Department of Astrophysical \& Planetary Sciences, University of Colorado, Boulder, CO 80309, USA}

  
\begin{abstract}
  Using cosmological simulations with a dynamic range in excess of
  $10^7$, we study the transport of gas mass and angular momentum
  through the circumnuclear region of a disk galaxy containing a
  supermassive black hole (SMBH). The simulations follow fueling over
  relatively quiescent phases of the galaxy's evolution (no mergers)
  and without feedback from active galactic nuclei (AGNs), as part of
  the first stage of using state-of-the-art, high-resolution
  cosmological simulations to model galaxy and black hole
  co-evolution. We present results from simulations at different
  redshifts ($z=6$, $4$, and $3$) and three different black hole
  masses ($3\times10^7$, $9\times10^7$, and $3\times10^8
  \dim{M}_{\sun}$; at $z=4$), as well as a simulation including a
  prescription that approximates optically thick cooling in the
  densest regions. The interior gas mass throughout the circumnuclear
  disk shows transient and chaotic behavior as a function of time. The
  Fourier transform of the interior gas mass follows a power law with
  slope $-1$ throughout the region, indicating that, in the absence of
  the effects of galaxy mergers and AGN feedback, mass fluctuations
  are stochastic with no preferred timescale for accretion over the
  duration of each simulation ($\sim 1$-$2 \dim{Myr}$). The angular
  momentum of the gas disk changes direction relative to the disk on
  kiloparsec scales over timescales less than $1 \dim{Myr}$,
  reflecting the chaotic and transient gas dynamics of the
  circumnuclear region. Infalling clumps of gas, which are driven
  inward as a result of the dynamical state of the circumnuclear disk,
  may play an important role in determining the spin evolution of an
  SMBH, as has been suggested in stochastic accretion scenarios.
\end{abstract}
  


\section{Introduction}

Supermassive black holes (SMBHs) with masses from $\sim 10^6$ to more
than $\sim 10^9 \dim{M}_{\sun}$ are found in the centers of most
galaxies \citep[e.g.,][]{KormRich, Magorrian98}. Observations of
correlations between the masses of SMBHs and several properties of
their host galaxies indicate that the growth of SMBHs is closely tied
to the evolution of their hosts \citep[e.g.,][]{Magorrian98, FM00,
  Geb00, Trem02}. Scenarios including feedback from active galactic
nuclei (AGNs) have been explored in order to understand both the
growth of SMBHs and the nature of their relationship with their host
galaxies \citep[e.g.,][]{SilkRees98, KaufHae00, WyitheLoeb03,
  DiMatteo05, Springeletal05a, Springeletal05b, Crotonetal06,
  DiMatteo08, Hopkins07, JohanssonNB09}. AGNs are thought to be
powered by accreting SMBHs \citep[e.g.,][]{LyndenBell69}, so that
observations of the various phases of AGN activity may provide clues
about the connection between SMBH growth and galaxy evolution, as
models explored by, e.g., \citet {Hopkins05} and \citet{Sijackietal07}
have shown. It is therefore essential to characterize accretion onto
SMBHs in order to understand AGNs and their relevance for galaxy
evolution.

Substantial SMBH growth requires a mechanism for replenishing fuel on
the subparsec scales feeding the SMBH accretion disk. Large scale
tidal torques caused by galaxy mergers are an effective mechanism for
transporting gas from super-galactic and galactic scales down to
scales of several parsecs, where other mechanisms may become more
viable for funneling material the rest of the distance toward the SMBH
\citep[e.g.,][]{Hernquist89, BarnesHernquist92}. Secular evolution may
also play a role as disk instabilities lead to the formation of bars
and spiral waves which can transport material \citep{Robertsetal79,
  barswinbars, RegTeu04}. There are several observations of inflowing
gas in the circumnuclear regions of low-luminosity AGNs, apparently
driven by global instabilities arising from secular evolution
\citep[see, e.g.,][]{NUGA4, NUGA7, StorchiBergmann07, NUGA8,
  Riffeletal08}. Mergers are thought to be effective for building
SMBHs in more massive early-type galaxies, which appear to form
earlier \citep[see, e.g.,][and references therein]{Hopkinsetal08a,
  Hopkinsetal08b}, ultimately building the SMBH-bulge relationships,
whereas secular evolution may play a larger role in the growth of
late-type galaxies, particularly at lower redshift \citep[see,
  e.g.,][and references therein]{KormKenn04}.

Whichever is the dominant mechanism for fueling SMBHs, the specific
details of how fueling occurs in the circumnuclear regions of galaxies
are still not understood. Fueling may occur continuously, over the
course of large-scale dynamical instabilities in the galaxy, or it may
be an intermittent process, dependent entirely on the dynamics on
small scales. Cold gas, stochastically accreted from the circumnuclear
regions of galaxies, may be able to sustain the fueling of
low-luminosity AGNs \citep{HopkinsHernquist06}. Intermittent accretion
episodes, consisting of in-falling clouds of gas with randomly
oriented angular momentum vectors, may contribute to the spin-down of
SMBHs, thus lowering their radiative efficiency and allowing them to
grow faster
\citep[see][]{KingPringle06,KingPringle07,Kingetal08}. \citet{Wangetal09}
have used a Soltan-type argument \citep{Soltan82} to show that the
average radiative efficiency of SMBHs has decreased over time, as
might be expected if SMBHs grow predominately through episodic
accretion events that act to lower their spin. Using cosmological
simulations which followed the buildup of large SMBHs at $z=6$,
\citet{Sijackietal09} showed that black hole mergers can contribute to
the spin-down of SMBHs as well, resulting in low radiative
efficiencies and rapid growth. A decrease in radiative efficiency must
be reconciled with observations of an increase in the fraction of
radio-loud quasars with decreasing redshift \citep{Jiangetal07}, which
would suggest SMBHs with larger spins (and hence larger efficiencies)
at low-$z$. Therefore, understanding the behavior of angular momentum
in the circumnuclear region is an important part of modeling
accretion, and subsequently AGN feedback.

Simulations that follow the growth of SMBHs over cosmological times,
or over the duration of a galaxy merger, often cannot follow the
circumnuclear regions of galaxies with high enough resolution to
describe the accretion flow in detail. Such simulations must make
approximations for the accretion rates using the properties of the
galaxies on scales that are resolved. A common technique is to assume
that the unresolved disk is fed by Bondi-type accretion
\citep{HoyleLyttleton39, BondiHoyle44, Bondi52}, as in for example the
smoothed particle hydrodynamics simulations of \citet{Springeletal05a}
who estimate the accretion rate based on the properties of the gas on
a scale of $\sim100 \dim{pc}$. The assumption of Bondi accretion
appears to be reasonable for following evolution over cosmological
times, where the average accretion rate onto the black hole cannot be
more than the average accretion rate through any radius, since the
fuel supply is limited by large scales. Previous cosmological
simulations have been successful in reproducing observed population
demographics and trends, such as the black hole mass function and
galaxy colors and morphologies \citep[as in, e.g.,][]{DiMatteo08,
  Croftetal09, McCarthyetal09}. However, for detailed studies of the
growth and evolution of individual SMBHs or of the coupling between
AGN feedback and the black hole accretion rate, more accurate
descriptions of the accretion rate and its dependence on the
small-scale features of the host galaxy are desirable.

Small-scale simulations have addressed gas dynamics in
subgalactic-scale disks with high resolution \citep{Fukuda00, Wada01,
  WadaNorman01,Escala07,WadaNorman07,KawakatuWada08}, finding the
development of a turbulent, multi-phase interstellar medium. The
approximation of Bondi accretion in such an environment is not
necessarily invalid. \citet{Krumholzetal05} have shown that modified
forms of the Bondi prescription can describe accretion in turbulent
environments. A simulation must be equipped to model the properties of
the turbulence in order to employ the modified Bondi prescription. If
the gas contains a significant amount of angular momentum, the
unmodified Bondi prescription gives inaccurate estimates of the
accretion rate as well \citep{ProgaBegelman03,
  Krumholzetal06}. However, even modified Bondi prescriptions become
inapplicable in the case of the self-gravitating, rotationally
supported disks that are likely to form as large amounts of gas are
driven inward in high-redshift galaxies. As the use of adaptive
techniques increasingly improves the resolution of cosmological
simulations, accretion onto black holes can no longer be described by
approximate prescriptions such as Bondi accretion.

In the present paper, we use cosmological adaptive mesh refinement
simulations with a large dynamic range to study the transport of gas
and angular momentum through the circumnuclear disk of an SMBH host
galaxy over time. The goal is to provide a description of accretion
that can be compared to prescriptions typically applied in larger
scale simulations. It will be shown that in the limiting case of
relative quiescence (e.g., in between merger events and without AGN
feedback) accretion is a stochastic rather than a continuous
process. The method behind the simulations used here is described in
detail in \citet{Levineetal08}, hereafter Paper I, and in
\citet{MyThesis}. In Section \ref{sec:simtr}, we briefly summarize the
details of the simulations. The results of the simulations are given
in Sections \ref{sec:mass}-\ref{sec:mom}, including an analysis of the
mass accretion rate and the angular momentum of the gas in the
circumnuclear region of the galaxy. Finally, we summarize and discuss
the results in Section \ref{sec:disctr}.


\section{SIMULATION}
\label{sec:simtr}

The simulations presented here were run using the Adaptive Refinement
Tree (ART) code \citep{Kravtsovetal97, KravtsovPhD, Kravtsovetal02},
following the ``zoom-in'' method described in detail in Paper I. The
code follows gas hydrodynamics on an adaptive mesh and includes dark
matter and stellar particles. The gas cooling and heating rates are
tabulated as functions of density, temperature, metallicity, and
redshift over the temperature range $10^2<T<10^9 \dim{K}$ using {\sc
  CLOUDY} \citep{Cloudy98}, which accounts for the metallicity of the
gas, the formation of molecular hydrogen and cosmic dust, and UV
heating due to cosmological ionizing background. Because the
circumnuclear disk of the simulated galaxy reaches high densities, the
gas can become optically thick to its own cooling radiation. In order
to account for this effect and approximate optically thick cooling, we
have also run a simulation that uses a Sobolev-like approximation for
the column density of the gas to account for the large opacity of the
highest-density regions in the simulated galaxy in calculating the
cooling rates \citep[see][]{Gnedinetal09}.

  \begin{deluxetable*}{lccccccc}
    \tabletypesize{\footnotesize}
    \tablecolumns{7}
    \tablewidth{\linewidth}
    \tablecaption{\label{tbl:run} Summary of Each Simulation Run}
    \tablehead{
      \colhead{Run} & \colhead{$z$} & \colhead{$M_{200}$} & 
      \colhead{$M_\textrm{g}$\tablenotemark{a}} & \colhead{$M_*$} & \colhead{$M_\bullet$} & 
      \colhead{$\Delta x_{\min}$} & \colhead{Duration} \\ 
      \colhead{}  &  \colhead{}  &  \colhead{($10^{11} h^{-1}\dim{M}_{\sun}$)} & 
      \colhead{($10^{10} h^{-1}\dim{M}_{\sun}$)} & \colhead{($10^9 h^{-1}\dim{M}_{\sun}$)} &
      \colhead{($10^7 \dim{M}_{\sun}$)} & \colhead{(proper pc)} & \colhead{($10^6 \dim{yr}$)}
    } 
    \startdata 
    Z3L20   & $3$  & $2.4$ & $3.3$ & $7.3$ & $3$  & $0.032$ & $0.84$ \\
    Z3.5L20   & $3.5$& $2.0$ & $2.8$ & $5.3$ & $3$  & $0.028$ & $0.71$ \\
    Z4L15   & $4$  & $1.7$ & $2.5$ & $4.0$ & $3$  & $0.819$ & $1.0$ \\
    Z4L18   & $4$  & $1.7$ & $2.5$ & $4.0$ & $3$  & $0.102$ & $1.0$ \\
    Z4L20\tablenotemark{b}  & $4$ & $1.7$ & $2.5$ & $4.0$ & $3$ & $0.026$ & $1.9$ \\
    Z4L20.B3   & $4$  & $1.7$ & $2.5$ & $4.0$ & $9$  & $0.026$ & $1.0$ \\
    Z4L20.B10   & $4$  & $1.7$ & $2.5$ & $4.0$ & $30$ & $0.026$ & $0.80$ \\
    Z4L20.OT\tablenotemark{c}  & $4$  & $1.7$ & $2.5$ & $4.0$ & $3$ & $0.026$ & $0.85$ \\
    Z6L20   & $6$  & $0.66$ & $1.0$ & $0.84$ & $3$ & $0.018$ & $1.9$ 
     \enddata
     \tablenotetext{a}{Mass of gas with $T<20{,}000 \dim{K}$.}
     \tablenotetext{b}{Fiducial run.}
     \tablenotetext{c}{Includes a Sobolev-like approximation for the column density to model optically thick cooling.}
  \end{deluxetable*}

Stellar particles are formed in cells with appropriate densities and
temperatures at an efficiency that matches observed star formation
rates on kiloparsec scales \citep{Kennicutt98} and $100 \dim{pc}$
scales \citep[e.g.,][]{Youngetal96, WongBlitz02}. Presently, only the
cosmological portion of the simulation (before zooming-in) includes
radiative transfer and feedback and enrichment from stars (although
stellar particles are still formed in the zoom-in simulations). The
physics on small scales in galaxies is complex, so our approach is to
individually address each problem, carefully building a realistic,
state-of-the-art simulation one piece at a time.

Here, we provide a brief overview of the zoom-in method, with a
complete description given in Paper I. A cosmological simulation is
evolved from a random realization of a Gaussian density field at
$z=50$ in a periodic box of $6 \chimps$ with an appropriate power
spectrum, and is followed assuming a flat $\Lambda$CDM model:
$\Omega_0=1-\Omega_{\Lambda}=0.3$, $\Omega_{\textrm{b}} = 0.043$,
$h=H_0/100=0.7$, $n_{\textrm{s}}=1$, and $\sigma_8=0.9$. Beginning
with the cosmological simulation, which has a maximum resolution of
$\approx 180 h^{-1} \dim{pc}$ ($37 h^{-1} \dim{pc}$ proper at $z=4$),
the resolution is slowly increased one refinement level at a time,
reaching a quasi-stationary state on each level before increasing to
the next level. The final, maximum resolution in our fiducial run is
$\approx 0.089 h^{-1} \dim{pc}$ ($0.02 h^{-1} \dim{pc}$ proper at
$z=4$), corresponding to $20$ levels of refinement.

After reaching the maximum resolution, a fraction of the gas in the
center of the galaxy is replaced with a black hole particle of equal
mass and momentum. The circumnuclear region is extremely gas rich,
with the gas comprising $91\%$, $83\%$, and $79\%$ of the baryon mass
inside a radius of $1 \dim{kpc}$ at $z=6$, $4$, and $3$, respectively.
In this region, the gas mass dominates over all other components,
including the black hole particle. The black hole mass has not been
chosen to match any of the local scaling relations
\citep[e.g.,][]{Magorrian98, FM00, Geb00, Trem02}.  Instead, we have
chosen a mass that is large enough to distinguish it from stellar
particles in the simulation, and that tends to stay put in the center
of the galaxy, but does not depend on any a priori assumptions about
how the scaling relations extend to high redshift.  The possibility
remains that the simulated galaxy should host a more massive black
hole, but since we do not follow the black hole's growth from its seed
formation, the mass of the black hole particle is somewhat arbitrarily
chosen. Presently, the mass of the black hole particle does not change
over the course of the zoom-in simulations. This simplification allows
us to avoid making any assumptions about how the gas mass on the
smallest scales of the simulations will be accreted onto the black
hole particle.

After the introduction of the black hole particle, the simulation
continues to evolve with the maximum resolution for several hundred
thousand years. The duration of each simulation varies with
redshift. At lower redshifts, the increased mass of the galaxy results
in more high-density cells, requiring high resolution. Following a
larger portion of the simulation with finer resolution is more
computationally expensive, therefore the lower redshift simulations
take longer to evolve. As a result, e.g., the $z=3$ zoom-in is evolved
only half as long in physical time as the fiducial $z=4$ simulation.

Most of the following analysis is applied to the short-term evolution
of the galaxy ($\sim 10^6 \dim{yr}$) during a zoom-in episode at $z=4$
containing a $3\times10^7 \dim{M}_{\sun}$ SMBH particle (considered
the fiducial run\footnote{The fiducial run, Z4L20, is effectively the
  same simulation described in Paper I. The only difference is in the
  initial zoom-in of the run presented here, which occurred more
  slowly (see discussion in Section 2.2, Paper I).}). Additional
simulations were performed, including zoom-in episodes at other
redshifts with the same black hole mass ($z=3$, $3.5$, and $6$), runs
with different black hole masses ($9\times10^7$ and $3\times10^8
\dim{M}_{\sun}$) at $z=4$, two lower resolution versions of the
fiducial run (with maximum levels of refinement $15$ and $18$), and a
version of the fiducial run that implements an approximation for
optically thick cooling. The $z=4$ simulation was chosen for the
fiducial run because this redshift corresponds to a relatively
quiescent stage in the growth of the host galaxy. At this stage, the
main galaxy of the simulation has not undergone a large mass ratio
merger in $\approx 400 \dim{Myr}$, although the galaxy is still
actively growing. At $z=6$ (also studied here), the galaxy is less
relaxed, as it is entering into a $4{:}1$ merger at that time. It is
expected that the mass of the black hole should continue to grow with
decreasing redshift. Since the different redshift simulations each
contain the same black hole particle mass, they do not provide the
growth history of a single galaxy, but rather they represent the
distinct growth histories of similar disk galaxies. Table
\ref{tbl:run} summarizes the details of the simulations performed,
including redshift, the total mass within a radius enclosing a mean
density of $200$ times the present-day critical density of the
universe, centered on the main galaxy ($M_{200}$), the mass of gas
with temperature less than $20{,}000 \dim{K}$ ($M_{\textrm{g}}$), the
stellar mass ($M_*$), the black hole mass ($M_{\bullet}$), the size of
the most refined cell in each run ($\Delta x_{\min}$), and the
duration of each run in physical time.

Paper I gives the details of the spatial structure of the
circumnuclear region for the fiducial run. Each run described here
develops a similar structure: a cold, rotationally supported,
self-gravitating gas disk develops in the circumnuclear region of the
galaxy (inside $\sim 100 \dim{pc}$), with a steep power-law density
distribution. The disk is globally unstable, leading to the
development of waves and instabilities, which drive turbulence on a
range of scales, injecting energy through shocks. The turbulence is
highly supersonic, effectively raising $Q$ and acting to maintain
local stability in the disk. In the run Z4L20.OT, the approximation
for optically thick cooling accounts for the densest gas being
optically thick to its own cooling radiation. Consequently, the
mass-weighted mean temperature in that run is higher than in the
fiducial run, which contributes to a slightly higher Toomre
$Q$-parameter. In either case, the higher effective $Q$ prevents the
circumnuclear disk from fragmenting entirely into star-forming clumps
on a free-fall time. However, the actual star formation rate in the
simulation does not reflect the complicated behavior of the
circumnuclear region, because the rate is simply density- and
temperature-dependent and operates on timescales much longer than the
million-year time scales of the zoom-in simulation. During the zoom-in
simulation, the disk shows transient features caused by spiral waves
and global instabilities, which slowly allow angular momentum
transport, driving gas inward toward the center of the galaxy. Similar
behavior, where global instabilities generate stabilizing turbulence,
has been observed in other numerical simulations
\citep{ReganHaehnelt09, WiseII, EnglmaierShlosman04} and has been
studied analytically by \citet{BegelmanShlosman09}. In the following
sections, we explore the behavior over time of the mass accretion rate
and the angular momentum of the gas throughout the circumnuclear disk.

\section{Disk Mass and the Rate of Transport}
\label{sec:mass}

  \begin{figure}[!t]
    \center{\includegraphics[width=.95\linewidth]{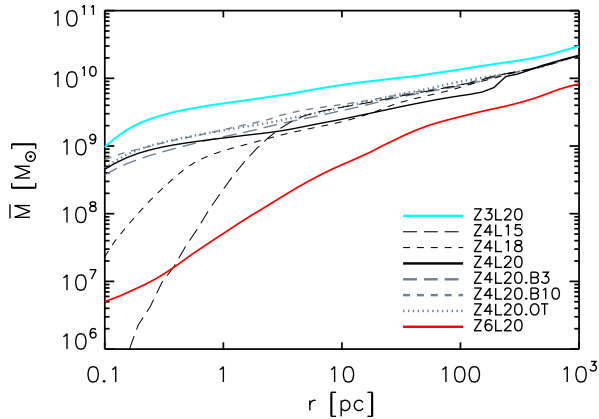}}
      \caption{\label{fig:gmr} Time-averaged interior gas mass as a
        function of radius for each simulation.}
  \end{figure}

Figure \ref{fig:gmr} shows the mean gas mass, $\mean{M}$ (averaged
over the duration of the simulation), interior to radius, $r$, for
each simulation (except $z=3.5$). The radius $r$ is measured with
respect to the position of the black hole particle, which approximates
the center of the galaxy over the duration of each zoom-in simulation
(except for a few brief intervals, which are discussed later in this
section). A comparison of the average gas mass in each zoom-in episode
approximately shows the growth of the circumnuclear disk over
cosmological times. The substantial growth in mass between $z=6$ and
$z=4$ is caused by a major merger (with a mass ratio of $4{:}1$) and a
few minor mergers occurring in between.

    \begin{figure*}
      \center{\includegraphics[width=.8\linewidth]{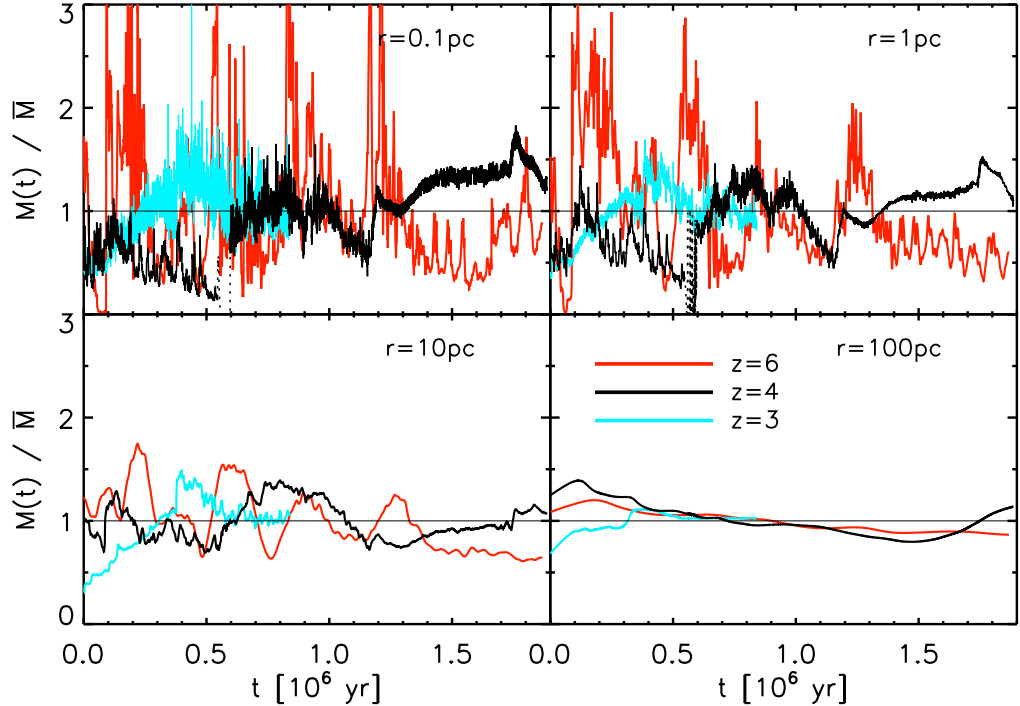}}
      \caption{\label{fig:gmz} Amplitude of gas mass fluctuations at
        redshifts $3$, $4$, and $6$ for four different radii. The
        dotted black line shows the fluctuations measured with respect
        to the black hole particle position over the duration of the
        black hole displacement. The thin solid line shows $M =
        \mean{M}$ for comparison. (Figure \ref{fig:gmr} shows
        $\mean{M}$ as a function of radius, $r$, for each
        simulation).}
  \end{figure*}

In contrast with the averaged mass-profiles, the instantaneous
mass-profiles at individual times during the zoom-in part of the
simulation (not shown) reveal the transient nature of the
circumnuclear disk. As the globally unstable disk forms transient
structures on a range of scales, the interior gas mass can vary
significantly. Figure \ref{fig:gmz} shows the amplitude of mass
fluctuations, $M(t)/\mean{M}$ at four different radii (ranging from
$0.1$ to $100 \dim{pc}$, spanning scales which have undergone at least
one rotation period over the course of each of the zoom-in
simulations) over time for each of the three redshift simulations. The
figure gives an indication of the turbulent nature of the gas in the
circumnuclear disk of the galaxy. In their high-resolution
re-simulations of mergers between disk galaxies,
\citet{HopkinsQuataert09} find a qualitatively similar picture, in
which global instabilities operate on a range of scales, causing a
great deal of fluctuation in the interior gas mass as a function of
time throughout the circumnuclear disk.

Around $550{,}000 \dim{yr}$ in the $z=4$ simulation, a $\sim$ few
$\times 10^7 \dim{M}_\sun$ clump of gas (which forms in the central
$10 \dim{pc}$, perhaps as a result of the global instabilities in the
disk) moves within a few parsecs of the black hole particle,
temporarily creating a gravitational disturbance that displaces the
black hole particle for $\approx50{,}000 \dim{yr}$. The black hole
soon resettles into the bottom of the potential well of the disk, but
because the interior mass is measured with respect to the black hole
particle, the interior gas mass measurements are temporarily affected
by the black hole's motion (see the dotted lines in Figure
\ref{fig:gmz}). Therefore, the interior gas mass shown in Figure
\ref{fig:gmz} has been interpolated over the duration of the black
hole displacement in all subsequent computations (unless otherwise
noted). The properties of the clump of gas may be significant for the
angular momentum of the black hole accretion disk, which will be
discussed further in Section \ref{sec:mom}. Runs Z4L20.BH3,
Z4L20.BH10, and Z4L20.OT undergo similar events of variable
duration. The details of each displacement, such as the size and the
duration of the subsequent relaxation, depend on the somewhat
arbitrarily chosen black hole particle mass and the gas dynamics,
which vary between different runs. Therefore, even though the effect
of in-falling clumps on the gas disk may be significant, the
displacement of the black hole particle in these simulations does not
necessarily imply that such displacements occur similarly in nature.

As Figure \ref{fig:gmz} shows, a characteristic accretion rate is not
straightforwardly determined by any one individual snapshot of the
circumnuclear region, given the amount of fluctuation in the gas mass
over the timescale of the simulations. A simple estimate of the net
accretion rate could be obtained by measuring the difference in mass
at each scale between the beginning and the end of each run. However,
the dynamical and turbulent behavior of the circumnuclear disk gives
rise to a highly variable accretion rate, often indicating as much
outflow as inflow of gas mass, and in some instances, resulting in a
net outflow. In this case, a Fourier transform of the interior gas
mass provides a natural characterization of the mass fluctuations and
their dependence (if any) on timescale.

We have chosen the sampling frequency for the mass (and all other
physical quantities), somewhat arbitrarily, to correspond to one time
step on level $6$. The frequency is such that the entire zoom-in
region in each of the simulations steps at least $8$ times
(corresponding to the least-resolved level of refinement in the
circumnuclear disk; level $9$), and at most $16{,}384$ times
(corresponding to the most-resolved level; level $20$)\footnote{Time
  steps forward in the simulation so that for each step on level $6$,
  the simulation has taken $2^{l-6}$ steps on level $l$.}. During the
simulation, the physical time corresponding to each step is
redetermined after every level-$0$ step, and depends on the dynamics
of the disk (so that the time step obeys a global CFL
condition). Therefore, the length of time between samplings varies
throughout the duration of the simulation. The Lomb-normalized
periodogram \citep[see][Numerical Recipes, Section $13.8$]{Lomb76,
  Press92}, or the spectral power normalized by the variance of a data
set, is useful for studying the behavior of unevenly sampled data,
such as the uneven time steps here. The normalized spectral power as a
function of the angular frequency, $\omega$, is given by

  \begin{align}\label{eq:lpow}
    P_{\textrm{N}}(\omega) & =
    \frac{\left[\sum\limits_j(M_j-\mean{M})\ \cos\omega(t_j-\tau)\right]^2}{2\sigma_{\textrm{M}}^2\ \sum\limits_j \cos^2\omega(t_j-\tau)} \nonumber \\ 
    & + \frac{\left[\sum\limits_j (M_j-\mean{M})\ \sin\omega(t_j-\tau)\right]^2}{2\sigma_{\textrm{M}}^2\ \sum\limits_j \sin^2\omega(t_j-\tau)},
  \end{align}

\noindent summed over $N$ data points, where $\sigma_{\textrm{M}}^2$
is the variance in the interior gas mass. The quantity $\tau$, defined
by

  \begin{equation}
    \tan(2\omega\tau) = \frac{\sum\limits_{j=1}^N \sin 2\omega
    t_j}{\sum\limits_{j=1}^N \cos 2\omega t_j},
  \end{equation}

\noindent is introduced so the spectral power is independent of
shifting the data in time. Although the Lomb-normalized periodogram is
not directly computed from the Fourier transform of the data, we use
it to estimate the Fourier transform of the interior gas mass as
follows:
 
   \begin{figure*}[!ht]
    \center{\includegraphics[width=0.8\linewidth]{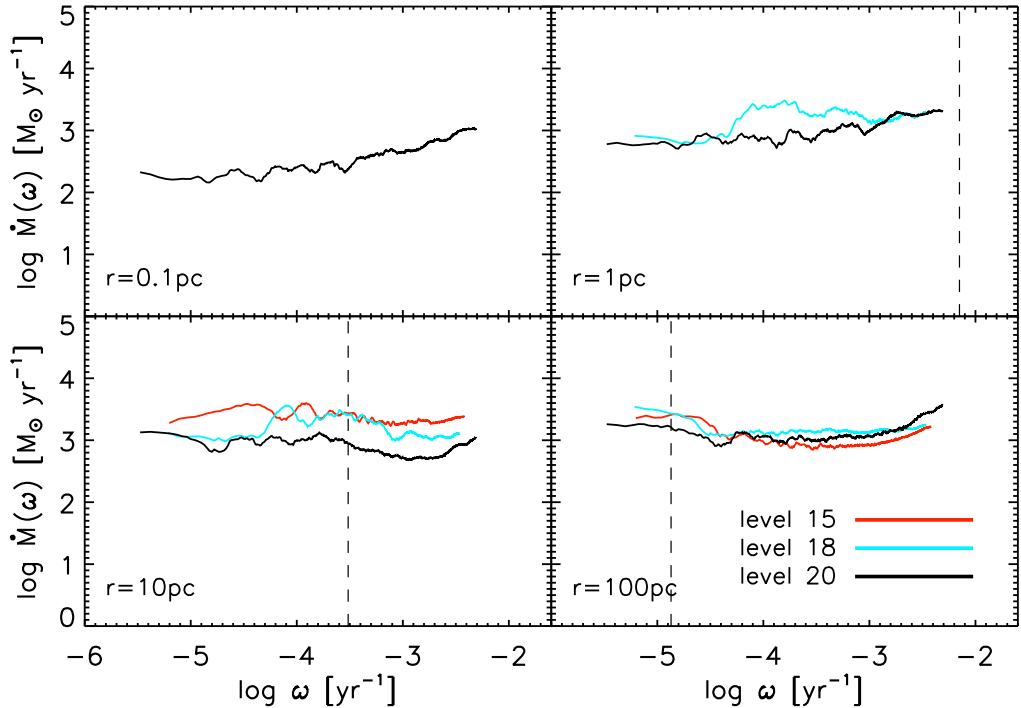}}
    \caption{\label{fig:mdotres} Mass flux, estimated from the
      spectral power of the interior gas mass fluctuations, for three
      different maximum resolution simulations ($15$, $18$, and $20$
      levels of refinement) at $z=4$. Here, the flux is only
      calculated for the first million years of run Z4L20, for a
      consistent comparison with runs Z4L15 and Z4L18.}
  \end{figure*}
 \begin{figure*}[!h]
    \center{\includegraphics[width=0.82\linewidth]{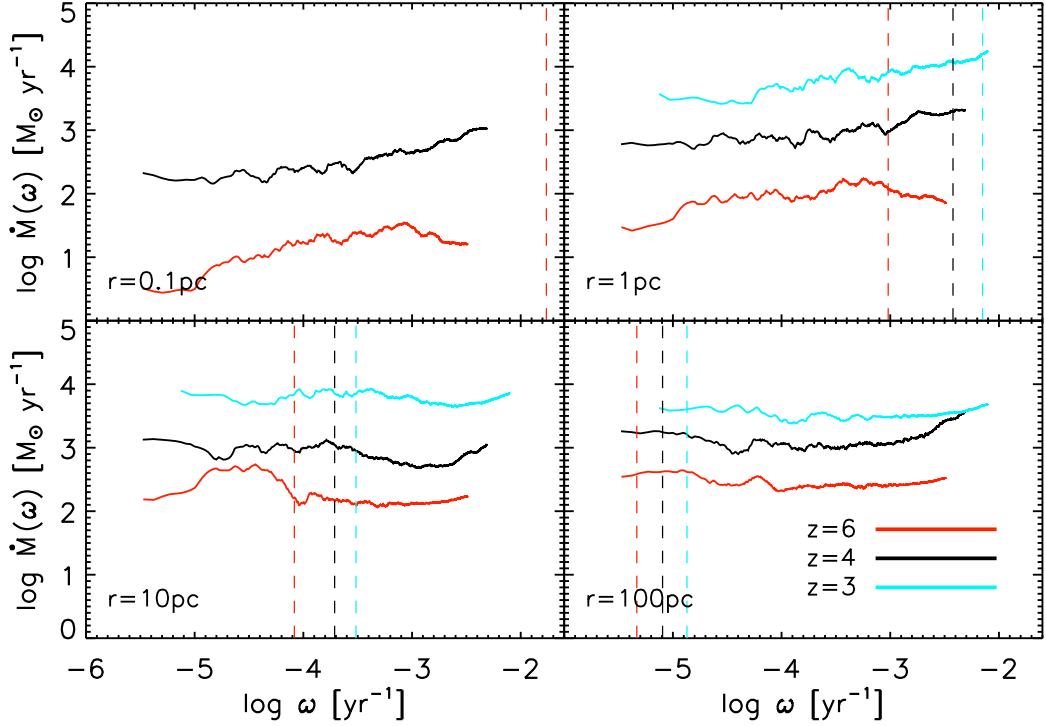}}
    \caption{\label{fig:mdotz} Mass flux, estimated from the spectral
      power of the interior gas mass fluctuations, for three different
      redshifts. The vertical lines correspond to the mean angular
      speed of the gas disk at radius, $r$, in each simulation.}
  \end{figure*}
   \begin{figure*}[!h]
    \center{\includegraphics[width=0.82\linewidth]{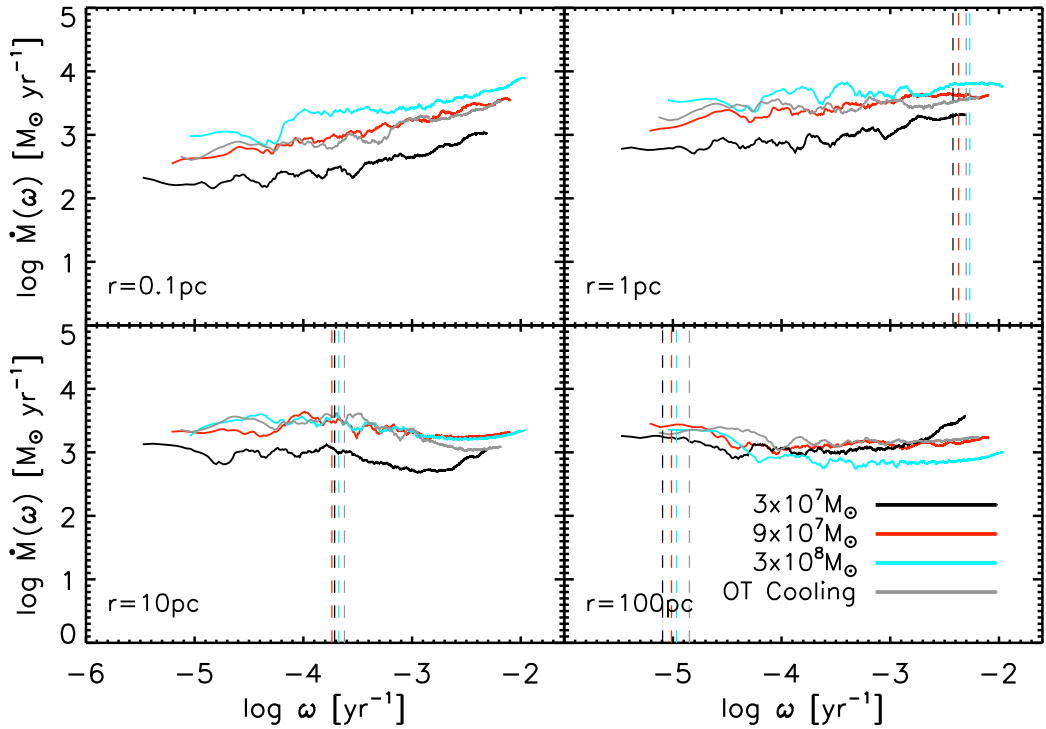}}
    \caption{\label{fig:mdotbh} Mass flux, estimated from the spectral
      power of the interior gas mass fluctuations, for three different
      black hole masses and for the optically thick cooling run at
      $z=4$. The vertical lines correspond to the mean angular speed
      of the gas disk at radius, $r$, in each simulation.}
  \end{figure*}

  \begin{align}\label{eq:fft}
    M(\omega) &= \frac{1}{N} \sum\limits_{j=1}^{N} M(t_j) e^{i \omega t_j} \nonumber \\
    &\approx \sqrt{\frac{P_{\textrm{N}}\sigma_{\textrm{M}}^2}{N}}.
  \end{align}

The Fourier transform of the mass follows a power law with slope
$\approx -1$ in each of the simulations, showing no substantial
departure down to frequencies comparable to the rotational period of
the disk. The time derivative of the mass Fourier transform gives a
characteristic ``accretion rate'' through radius $r$ as a function of
frequency,

  \begin{equation}\label{eq:fmdot}
    \dot{M}(\omega) = i \omega M(\omega).
  \end{equation}

\noindent Strictly speaking, the accretion rate given by Equation
\ref{eq:fmdot} does not describe the net inward transport of gas, but
rather it gives typical values of the flux of gas mass into and out of
a given region on a timescale $2\pi / \omega$. The absolute value of
the flux given by Equation \ref{eq:fmdot} is shown as a function of
frequency in Figures \ref{fig:mdotres}-\ref{fig:mdotbh}. The data are
smoothed over neighboring points (the number of neighbors used in the
smoothing is proportional to angular frequency, $\omega$) to reduce
the noise so that the different curves are readily
distinguishable. Figure \ref{fig:mdotres} shows a comparison of the
$z=4$ runs with different maximum resolution (Z4L15, Z4L18, and
Z4L20). The mass flux in each run is only shown for scales that are
resolved by four cells or more. We find good agreement in the behavior
shown for different resolution simulations, indicating that the
results are converged with respect to resolution.

The mass flux is larger for decreasing redshift, because the galaxy
itself is more massive, but the flux follows the same stochastic
behavior at different epochs (Figure\ref{fig:mdotz}), as well as for
different black hole masses and for the optically thick cooling run
(Figure \ref{fig:mdotbh}). We note that $0.1 \dim{pc}$ (proper) is the
resolution limit at $z=4$ (corresponding to four level-$20$ cells),
and therefore this scale is barely resolved at $z=4$ and
under-resolved at $z=3$. Therefore, the top-left panel of Figure
\ref{fig:mdotz} does not show the $z=3$ simulation. The vertical lines
in each panel correspond to the average angular speed of the gas at a
given scale in each simulation. The fluctuations do not appear to be
correlated with the orbital period at any radius in the circumnuclear
disk. The flat slope in the flux, $\dot{M}(\omega)$, shows that the
rate of transport is independent of frequency, i.e., that accretion is
a stochastic process with no preferred timescale. The result is
generic over each run studied here, including the optically thick
cooling run, which is dynamically similar to the other runs despite
the presence of higher temperature gas throughout the circumnuclear
disk. Simulations of self-gravitating disks show behavior similar to
our own results, with instabilities driving turbulence
\citep{WadaNorman99, Wada01, Wadaetal02}, which ultimately leads to
intermittent fueling episodes varying on timescales of $10^4-10^5
\dim{yr}$ \citep{Fukuda00, Wada04}.

\section{COMPARISON TO OTHER ACCRETION PRESCRIPTIONS}
\label{sec:acc}

As mentioned in the Introduction, simulations that do not resolve the
accretion disk of an SMBH, or the region where the gravity of the SMBH
dominates, typically include some prescription for the accretion rate
that depends on the properties of the simulated galaxy on scales that
are resolved. It is useful to check how well a prescription, such as
Bondi accretion, describes mass transport when extended down to parsec
scale resolution.

The original Bondi prescription assumes pressure-supported gas in
hydrodynamic equilibrium. Turbulence, as well as angular momentum
(such as in a disk), has been shown to lower the actual accretion rate
below the Bondi rate
\citep{AbramowiczZurek81,ProgaBegelman03,Krumholzetal06}. The
circumnuclear disk in the simulations here is both turbulent and
rotationally supported (see Figure 6 of Paper I), bearing little
resemblance to the medium modeled in the original Bondi
prescription. Modifications to the Bondi prescription that account for
both turbulence and vorticity in gas can be applied to estimate the
reduced accretion rate \citep{Krumholzetal05,Krumholzetal06}. However,
even such a modified prescription fails to describe accretion of
self-gravitating gas, such as in the circumnuclear disk of the present
simulations. Disks that are self-gravitating form bars and spiral
waves that further change the flow of gas.

\begin{figure}[!t]
   \center{\includegraphics[width=\columnwidth]{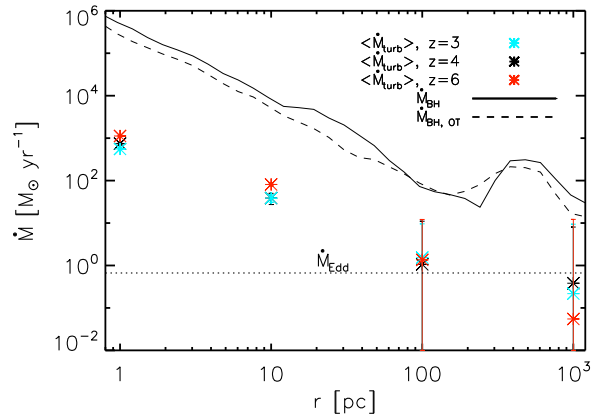}}
   \caption{\label{fig:mbondi} Bondi accretion rate as a function of
     radius for different redshifts. Two methods of estimation are
     shown: (1) $\dot{M}_{\textrm{BH}}$ computed from spherically
     averaged quantities (solid black curve: fiducial run, dashed
     curve: optically thick cooling run) and (2) the geometric mean
     and standard deviation of the $\dot{M}_{\textrm{turb}}$ PDF
     (points). The Eddington accretion rate, $\dot{M}_{\textrm{Edd}}$
     for a $30$ million $\dim{M}_{\sun}$ black hole with a radiative
     efficiency of $0.1$ is also shown for comparison (dotted black
     line).}
 \end{figure}

Figure \ref{fig:mbondi} shows a comparison of the Bondi Hoyle rate for
turbulent gas (computed using the spherically averaged radial profiles
of the gas density, turbulent velocity, and sound speed) and the
vorticity-reduced rate, applied to each redshift simulation. In
calculating the Bondi Hoyle rate for turbulent gas,
$\dot{M}_{\textrm{BH}}$, the thermal sound speed of the gas has been
replaced by an effective sound speed defined as the quadrature sum of
the thermal sound speed and turbulent velocity. The vorticity-reduced
rate, $\dot{M}_{\textrm{turb}}$, given by Equation 3 of
\citet{Krumholzetal06}, is computed cell by cell, producing a
log-normal probability density function (PDF) of accretion rates
within individual spherical shells centered on the black hole
particle. The geometric mean of the distribution, $\langle
\dot{M}_{\textrm{turb}}\rangle$, gives characteristic accretion rates
under this prescription and is shown for several different radii in
Figure \ref{fig:mbondi}. The vorticity-reduced rates are $2-3$ orders
of magnitude smaller than the standard Bondi Hoyle rates. These
results are qualitatively consistent with those of
\citet{Debuhretal09}, who find that accretion in their merger
simulations is regulated by angular momentum transport processes in
the host galaxy, keeping the accretion rate below the Bondi rate,
except during peak activity (when the accretion rate approaches
$\dot{M}_{\textrm{Edd}}$, which is $\approx \dot{M}_{\textrm{BH}}$).

As revealed in Figure \ref{fig:mbondi}, the Bondi prescription gives
large estimates for the accretion rates ($\gg 10
\dot{M}_{\textrm{Edd}}$) inside the central few hundred parsecs. The
steep, power-law, density profile of the gas in the circumnuclear disk
contributes to the large accretion rates shown in Figure
\ref{fig:mbondi} (which are proportional to density). The inclusion of
an approximation for optically thick cooling in the simulations does
not significantly change the density profile or the velocity
dispersion of the gas. Therefore, the Bondi prescription produces
similar results for runs Z4L20 and Z4L20.OT (dashed and solid curves
in Figure \ref{fig:mbondi}). Additional effects not included in our
current simulations may contribute to the depletion of gas in the
circumnuclear disk, such as AGN feedback and stellar feedback (not
included in the zoom-in portion of the simulation), thus lowering the
estimated Bondi rate. However, if the disk remains self-gravitating
and therefore susceptible to instabilities, the Bondi prescription
will inaccurately describe the transport of gas through this region.

\begin{figure}[!t]
    \center{\includegraphics[width=\columnwidth]{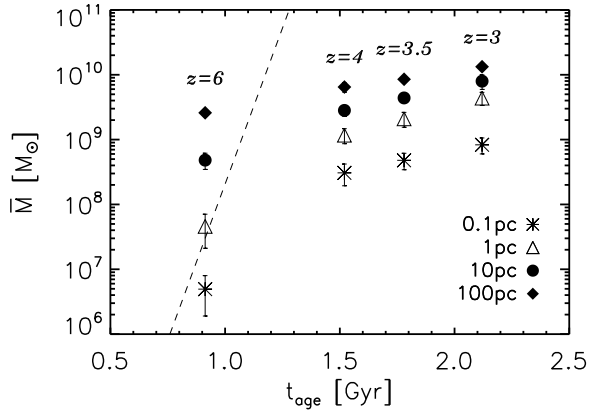}}
      \caption{\label{fig:mz} Mean gas mass interior to radius $r$ as
        a function of the age of the universe, $t_{\textrm{age}}$. The
        error bars correspond to the dispersion in the interior gas
        mass over each run. A $z=3.5$ simulation, not mentioned above,
        is also included in the figure.}
  \end{figure}

The average accretion rate through the circumnuclear disk over
cosmological times can be estimated by comparing the mean interior gas
mass (the averages shown in Figure \ref{fig:gmr}) for the different
redshift simulations. Figure \ref{fig:mz} shows the mean gas mass,
interior to radius $r$ as a function of the age of the universe,
$t_{\textrm{age}}$. Once again we emphasize that the different
redshift simulations do not necessarily describe different stages of
growth of the same galaxy because they each contain the same mass SMBH
(rather than a black hole that grows with redshift). However, the
black hole particle does not currently play a large role in the
evolution of the simulated galaxy (at least not below $z=4$, where the
mass of the black hole is dominated by the gas mass all the way down
to the resolution limit).

The dashed line in Figure \ref{fig:mz} shows the mass of a black hole,
initially $3\times10^7 \dim{M}_{\sun}$ at $z=6$, if it grows
continuously at the Eddington limit according to

  \begin{equation}\label{eq:sal}
    M(t) = M_{\bullet}\ \textrm{e}^{t/t_{\textrm{S}}},
  \end{equation}

\noindent where $t_{\textrm{S}}$ is the Salpeter time
\citep{Salpeter64} of $4.5\times10^7 \dim{yr}$, for a radiative
efficiency $\eta=0.1$. If the dynamics of the circumnuclear disk
extend all the way down to scales beneath the resolution (which cannot
be assumed) then the growth rate approximated by the simulation
points, which appears to steepen at early times with decreasing scale,
may continue to steepen down to smaller scales, approaching the
Eddington limit. Therefore, a black hole in our simulations will be
able to grow according to Equation \ref{eq:sal} at high-$z$, as long
as efficient fueling from the circumnuclear disk can be
sustained. This is perhaps consistent with the isolated disk galaxy
simulations of \citet{Springeletal05a}, where a sufficiently massive
black hole will grow according to Equation \ref{eq:sal}, in the
absence of AGN feedback. Once gas reaches sub-parsec scales, the
actual accretion rate onto the black hole is governed by the physics
of the accretion disk, not modeled by our simulations. Nonetheless,
Figure \ref{fig:mz} shows that the increase in the mass of the
circumnuclear gas disk over cosmological times yields enough fuel to
build up the mass of the SMBH if the fuel can be efficiently accreted.


\section{ANGULAR MOMENTUM}
\label{sec:mom}

The large fluctuations in the interior gas mass correspond to waves
moving through the circumnuclear disk, which can potentially drive
individual clumps of molecular gas into the vicinity of the SMBH. The
fluctuations develop stochastically, as shown in Section
\ref{sec:mass}, which may ultimately lead to stochastic SMBH accretion
events. If the direction of the angular momentum vector of accreted
material with respect to the black hole's spin axis also varies
stochastically, some fraction of accretion events will be
anti-aligned, lowering the black hole's spin and consequently its
radiative efficiency
\citep{KingPringle06,KingPringle07,NayakshinKing07,Kingetal08}. However,
large amounts of accretion, as in merger driven fueling, will tend to
align an SMBH's spin with that of the circumnuclear disk, increasing
its radiative efficiency \citep{Volonterietal07}. Spin-down resulting
from accretion could be more effective in non-merging disk galaxies
hosting smaller SMBHs, as secular processes stochastically drive small
amounts of gas toward the SMBH. The resulting low spin may be
consistent with observations of disk galaxies as hosts for radio-quiet
sources, if radio-loudness is associated with the spin of the black
hole \citep{BertiVolonteri08}. Since black hole mergers may also
contribute to the spin-down of SMBHs \citep{Sijackietal09}, it is
important to qualify the relative importance of gas accretion for
determining black hole spin.

The simulations here do not model the gas down to scales inside the
accretion disk, nor allow the black hole particle to accrete material
from larger scales (as discussed in Section \ref{sec:simtr}). For
these reasons, the simulations do not follow the spin of the black
hole. The simulations do model gas on parsec scales, enabling us to
resolve the direction of the angular momentum vector of gas and to
provide boundary conditions for simulations of SMBH accretion disks.

The angular momentum vector is measured with respect to the black hole
particle's position and velocity. Some uncertainty in the direction of
the angular momentum vector could arise because the position of the
black hole does not necessarily coincide exactly with the position of
the center of the galaxy and because the black hole may have a
significant rotational velocity about the center. During most of the
simulation, the black hole particle lies in the densest cell, which
coincides with the minimum of the galaxy's gravitational
potential. Therefore, the black hole particle typically provides a
reasonable estimate of the location of center of the galaxy. We have
also measured the angular momentum relative to a center whose velocity
is defined by an average over the central cell and six of its
neighbors, but find that the effect on the angular momentum vector is
negligible (except for the brief periods of black hole particle
displacement).
  
\begin{figure*}[!ht]
    \center{\includegraphics[width=.9\linewidth]{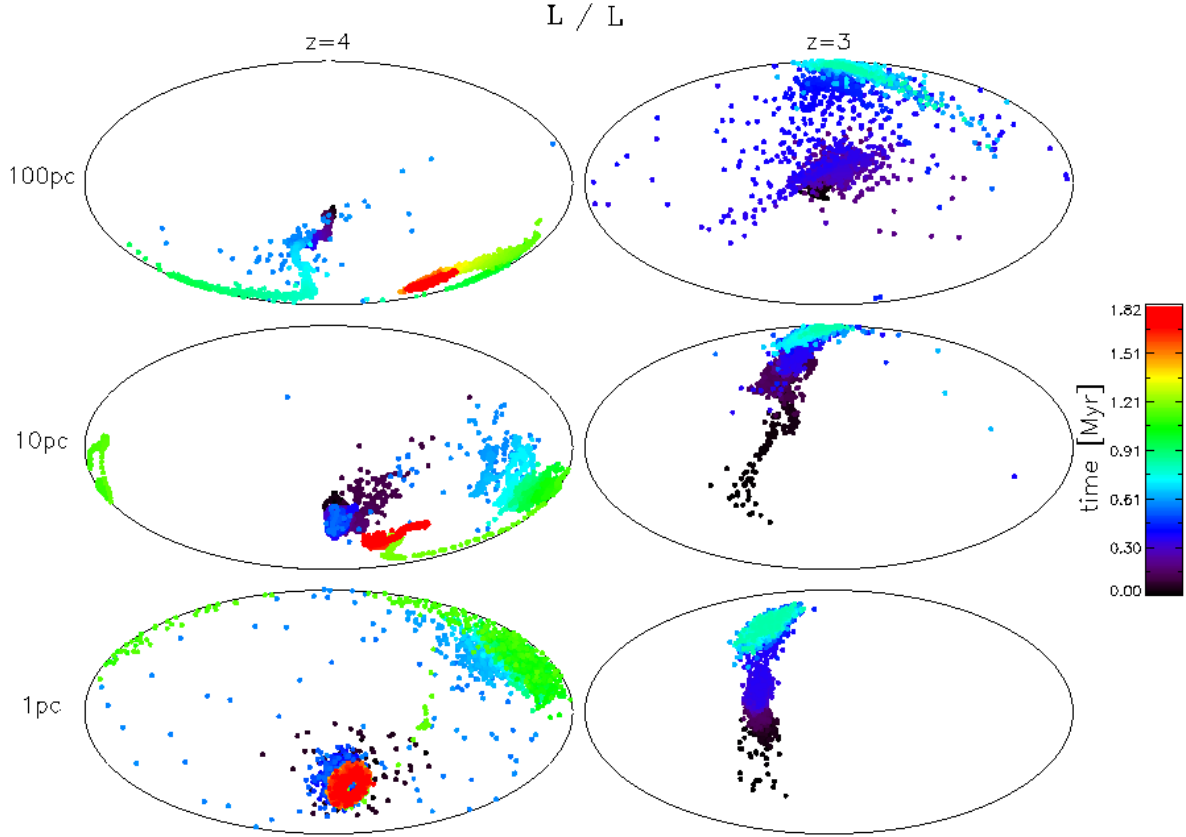}}
      \caption{\label{fig:jmap} Map showing the direction of the
        normalized angular momentum vector, $\boldsymbol{L}/L$ as it
        changes over time at $100$, $10$, and $1 \dim{pc}$ in the
        $z=4$ and $z=3$ simulations. The colors correspond to the time
        elapsed in the simulation since the introduction of the black
        hole particle. Each map is aligned so that the mean rotation
        axis of the galactic disk at a scale of $1 \dim{kpc}$ lies in
        the center. This figure is best viewed in color.}
  \end{figure*}

\begin{figure*}[!ht]
    \center{\includegraphics[width=0.77\linewidth]{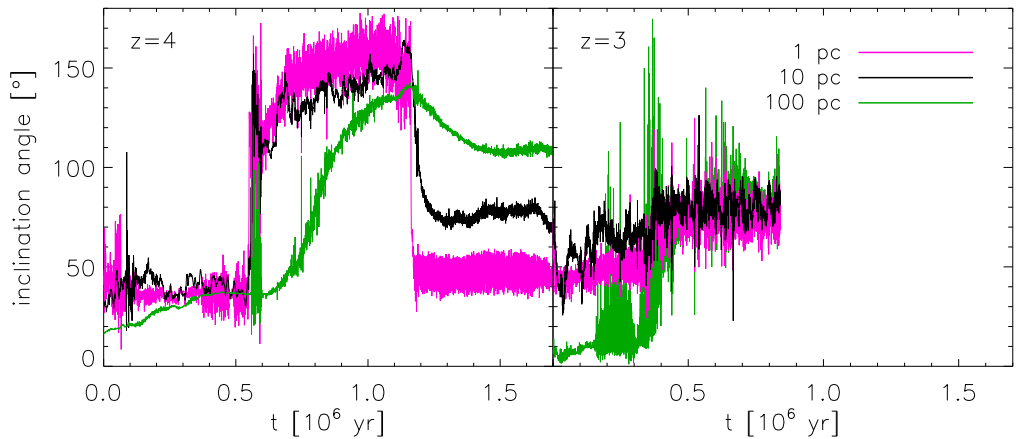}}
      \caption{\label{fig:thet} Inclination angle of the angular
        momentum of the inner disk with respect to that of the
        kiloparsec disk over physical time, for $z=4$ and $z=3$ (left
        and right panels, respectively).}
  \end{figure*}

Figure \ref{fig:jmap} shows a map projection of the direction of the
normalized angular momentum vector as it evolves in time at $100$,
$10$, and $1 \dim{pc}$ from the black hole particle for the $z=4$ and
$z=3$ simulations (left and right, respectively). The mean direction
of the rotation axis of the galaxy on kiloparsec scales, which remains
comparatively constant over the $\sim 1 \dim{Myr}$ duration of the
zoom-in simulation, is oriented toward the center of the map in each
case. The disk is slightly warped on scales smaller than a kiloparsec,
so that the axis of the circumnuclear disk is oriented at an angle to
the large-scale disk. Figure \ref{fig:thet} shows the inclination
angle of the angular momentum vectors shown in Figure \ref{fig:jmap}
with respect to those measured at a kiloparsec, over time. Together,
Figures \ref{fig:jmap} and \ref{fig:thet} demonstrate the effect of
the circumnuclear disk's chaotic behavior on the angular momentum of
the accreting gas.


At $z=4$ and a scale of $100 \dim{pc}$, the direction of the angular
momentum vector changes slowly but shows little scatter over the
course of the simulation. There is a small amount of scatter during
the displacement of the black hole particle, which is visible on each
scale shown. Rather than interpolating the magnitude and direction of
the angular momentum vector of the gas over the duration of the black
hole displacement, as was done with the interior gas mass profile in
Figure \ref{fig:gmz}, we continue to measure quantities relative to
the black hole particle. Once the black hole particle resettles into
the bottom of the potential well, the excess scatter in
$\boldsymbol{L}/L$ vanishes.  The slow change in direction at $100
\dim{pc}$ may correspond to the increased warping of the disk as the
simulation progresses. In a similar analysis of simulations of the
growth of a disk galaxy by \citet{SaitohWada04}, the spin axis of the
central gas disk also changed direction on $100 \dim{pc}$ scales
(relative to kiloparsec scales).
  
\begin{figure}[!t]
    \center{\includegraphics[width=.95\columnwidth]{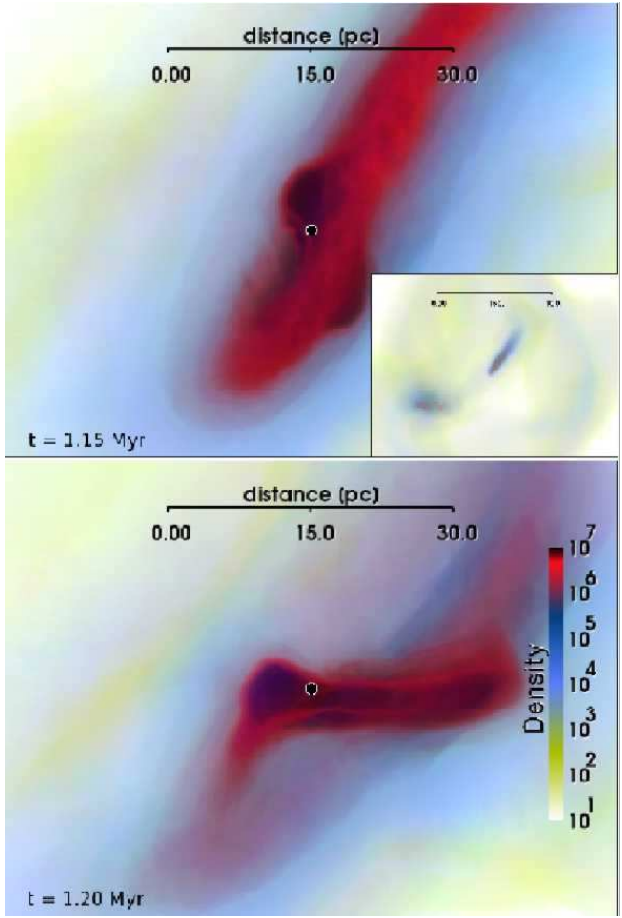}}
      \caption{\label{fig:flip} 3D volume-rendering of the gas density
        before ({\it top}) and after ({\it bottom}) the second angular
        momentum shift in the fiducial run referred to in the text
        (and in Figures \ref{fig:jmap} and \ref{fig:thet}). The inset
        in the top panel shows the large clump developing at a few
        hundred parsecs that is influencing the dynamics of the inner
        circumnuclear disk. The circle in the center of each panel
        corresponds to the position of the black hole particle.  This
        figure is best viewed in color.}
  \end{figure}

The angular momentum vector at $10$ and $1 \dim{pc}$ starts out with
the same approximate orientation as the larger scales but shows
slightly more scatter. There are two sudden changes in the direction
of the angular momentum vector at the $1 \dim{pc}$ scale. In the first
incident, at $\sim 0.55 \dim{Myr}$, the angular momentum changes
abruptly by more than $100 \degree$, coinciding with the time of the
displacement of the black hole particle described in Section
\ref{sec:mass} (as a clump of gas forms near and moves into the
center). The flip in direction is also visible at $10 \dim{pc}$. In
the second incident, at $\sim 1.2 \dim{Myr}$, the angular momentum
shifts by $\sim 100 \degree$ while the black hole remains stationary
at the center of the galaxy. The second shift is the result of
gravitational interaction with a $\sim$ few $\times 10^8
\dim{M}_\odot$ clump of gas, which develops at $> 100 \dim{pc}$ and
continues to move toward the center of the disk at late times. The
mass of the clump is comparable to the mass of the disk interior to
$\sim 10 \dim{pc}$. The second shift is also visible at the $10
\dim{pc}$ scale but occurs less suddenly. The two panels in Figure
\ref{fig:flip} show three-dimensional volume-renderings of the gas
density at $30 \dim{pc}$ before and after the flip (top and bottom,
respectively). The inset shows a zoomed-out view of the disk, where
the large clump of gas is visible $\gtrsim 200 \dim{pc}$ from the
center. If the clump remains intact as it reaches the center of the
circumnuclear region, a corresponding temporary displacement of the
black hole particle is expected, similar to the earlier
displacement. The black hole particle is likely to be displaced when
the in-falling gas mass is comparable to or greater than the black
hole mass and the mass of the gas disk interior to the in-falling
clump. However, a displacement has not yet occurred by the end of the
simulation. Similar angular momentum flips occur over the course of
runs Z4L20.BH3, Z4L20.BH10, and Z4L20.OT (not shown), often
corresponding to coincident displacements of the black hole particle.

The $z=3$ simulation shows more scatter at $100 \dim{pc}$ than the
$z=4$ simulation. There is a rapid shift in the direction of the
angular momentum vector of $\sim 30-45\degree$ beginning at $\sim 0.35
\dim{Myr}$ in the $z=3$ simulation. The shift is less dramatic than
both of those observed in the $z=4$ simulation. There is no
corresponding black hole particle displacement, likely because the
mass of in-falling clump is small compared to the mass of the disk
(and the black hole). The orientation of the disk appears to be more
coherent between the different scales in $z=3$ simulation than in the
$z=4$ simulation. This may be a result of the circumnuclear disk being
more massive at lower redshift. In broad terms however, the behavior
of the angular momentum vector in each of the simulations consistently
shows that the gas delivered to the SMBH may have varying angular
momentum, which can shift suddenly as massive clumps of gas develop
and move through the disk. The properties of the gas that is
ultimately accreted may then determine the spin and subsequently the
mode of feedback produced by the SMBH.


\section{DISCUSSION AND CONCLUSIONS}
\label{sec:disctr}

We have studied the transport of gas in the circumnuclear region of a
disk galaxy within a cosmological simulation at different redshifts,
for different SMBH masses, and for a run including an approximation of
optically thick cooling. The mass accretion rate is not steady, but
fluctuates randomly and substantially, the accretion rate through the
circumnuclear disk being almost as often negative as positive.  This
result is consistent with that of \citet{HopkinsQuataert09}, who find
a large amount of variation in the instantaneous accretion rates in
their galaxy simulations on scales $< 1 \dim{kpc}$. We find that
there is no preferred timescale for accretion over the course of the
$\sim$million years spanned by each of the simulations. The flat slope
of the Fourier transform of the accretion rate characterizes the
fluctuations as a function of timescale, revealing the stochastic
nature of accretion in the simulations. This complex and chaotic
behavior is apparent at each of the three redshifts explored in detail
here: $z=3$, $4$, and $6$, as well as in $z=4$ simulations containing
black holes with larger masses, and the $z=4$ run including optically
thick cooling.
 
The dynamic and turbulent nature of the simulated circumnuclear disk
is not well modeled by Bondi accretion, even when using a modified
prescription that takes into account the vorticity and turbulent
properties of the gas. Bondi prescriptions do not take into account
the effect of the disk's self-gravity.  Additionally, the modified
Bondi accretion rates, estimated from the properties of the disk in
the highest-resolution part of the simulation (where the density rises
steeply with decreasing radius), are highly super-Eddington on the
smallest scales (assuming a radiative efficiency of $0.1$). As the
Eddington rate provides an upper limit for the accretion rate in the
presence of radiative feedback from the SMBH, highly super-Eddington
rates are not likely to endure in nature. 

The Eddington rate (or a few times the Eddington rate) is set as an
upper limit to the accretion rate in many of the simulations that
employ Bondi type prescriptions to model accretion
\citep[e.g.,][]{Springeletal05a, DiMatteo05, Lietal07b, DiMatteo08}. On
scales corresponding to the spatial resolution of those simulations
($\gtrsim 100 \dim{pc}$), the accretion rates predicted by the
modified Bondi prescription in the present simulations are closer to
the Eddington limit than they are at smaller scales. However, black
holes in other simulations spend the majority of time accreting well
below this limit, at substantially sub-Eddington rates. The difference
between the rates determined here and those of other simulations is
even larger when comparing with prescriptions that multiply the Bondi
accretion rate by a large factor to approximate the effects of the
small scale physics, since no such factor has been included in our
calculations \citep[see discussion of the parameter
  $\alpha$][]{BoothSchaye09, JohanssonBN09, JohanssonNB09}. However,
it is not straightforward to make a direct comparison between
accretion rates in the present simulations and the rates in other
simulations that contain prescriptions for AGN feedback. As shown in
galaxy merger simulations \citep[e.g.,][]{Springeletal05a,
  Debuhretal09}, the presence of AGN feedback regulates black hole
growth by forcing gas out of the circumnuclear region. Future work
incorporating AGN feedback into the present simulations will address
this issue.

The absence of AGN feedback in our simulations may result in an
excessively dense circumnuclear gas disk, which corresponds to large,
super-Eddington accretion rates within the Bondi
prescription. Allowing the black hole particle to grow in situ in our
simulations, or using a more physically motivated recipe for star
formation in the zoom-in simulation would further change the
distribution of mass in the circumnuclear gas disk. The circumnuclear
disk is self-gravitating, almost certainly behaving outside the Bondi
regime, even with the modifications of \citet{Krumholzetal06}. This
result reinforces the need for high-resolution simulations
incorporating detailed gas dynamics and radiative processes, in order
to truly model the complicated dynamics on small scales, which
ultimately govern accretion onto SMBHs.

The chaotic behavior of the disk may be consistent with models
suggesting that stochastic accretion of molecular clouds from the
circumnuclear region of the galaxy can power low-luminosity AGNs
\citep{HopkinsHernquist06}. This manner of accretion may determine the
spin of the black hole. We find that the angular momentum vector on
scales $\lesssim 100 \dim{pc}$ can vary substantially from the
direction of angular momentum on kiloparsec scales. In each
simulation, the axis of the disk is closely aligned with the axis on
kiloparsec scales at the beginning of the zoom-in simulation, but
slowly over time shifts direction as the circumnuclear disk
dynamically evolves.

The angular momentum can be misaligned with the rest of the
circumnuclear disk, resembling the stochastic fueling scenarios
mentioned in Section \ref{sec:mom}. As is observed in several of the
simulations presented here, clumps of gas develop in the circumnuclear
disk, which change the angular momentum of the gas that ultimately
feeds the accretion disk. The change can occur as clumps fall into the
center or as larger clumps develop within the disk and change the
potential. However, as cautioned above, the simulation does not yet
include feedback from an accreting black hole, which will heat the gas
and drive out some of the material in the circumnuclear disk. Future
work including more realistic physics in the simulations will allow us
to constrain better the effects of accreting gas on the properties of
the SMBH and the feedback it produces.

\acknowledgements

The authors thank the anonymous referee for constructive comments and
suggestions. This work was supported in part by the DOE and the NASA
grant NAG 5-10842 at Fermilab and by the NSF grants AST-0134373,
AST-0507596, and AST-0708607. Supercomputer simulations were run on
the IBM P690 array at the National Center for Supercomputing
Applications and San Diego Supercomputing Center (under grant
AST-020018N), as well as on the Joint Fermilab-KICP Supercomputing
Cluster (supported by grants from Fermilab, the Kavli Institute for
Cosmological Physics, and the University of Chicago).


\bibliography{ms}

  
\end{document}